\documentclass[12pt]{article}

 \newread\testifexists
 \def\GetIfExists #1 {\immediate\openin\testifexists=#1
     \ifeof\testifexists\immediate\closein\testifexists\else
     \immediate\closein\testifexists\input #1\fi}

 \usepackage{gthstyle}\usepackage{amsfonts}
 \usepackage{amssymb}
 \usepackage{graphicx} \usepackage{epstopdf}
 \def\bbf#1{\setbox0=\hbox{$#1$} \kern-.025em\copy0\kern-\wd0
         \kern.05em\copy0\kern-\wd0 \kern-.025em\raise.0433em\box0}

 \immediate\openin\testifexists=a
     \ifeof\testifexists\immediate\closein\testifexists\else
     \immediate\closein\testifexists\input a\fimssym.def  

                     \newcommand{\fn}{\footnote}
              \newcommand{\nm}{\nonumber}
 \newcommand{\be}{\begin{eqnarray}}             \newcommand{\ee}{\end{eqnarray}}
 \newcommand{\bi}[1]{\begin{itemize}\item[#1]}         \newcommand{\itm}[1]{\item[#1]}  \newcommand{\ei}{\end{itemize}}
 \newcommand{\eqn}[1]{(\ref{#1})}

 \newcommand{\crlb}[1]{\label{#1}\\[2pt]}
 \newcommand{\eela}[1]{\quad\hbox{\scriptsize{#1}}\label{#1}\end{eqnarray}}
 \newcommand{\eelb}[1]{\label{#1}\end{eqnarray}}
 
 \newcommand{\newsecb}[2]{\section{#1}\label{#2}\setcounter{equation}{0}}

 \newcommand{\nolabels} {\def\eel{\eelb} \def\crl{\crlb} \def\newsecl{\newsecb}\def\bibiteml{\bibitem}\def\citel{\cite}   \def\lab{\label}}

\newcommand\publishversion{\nolabels\setlength{\textheight}{9in}\setlength{\oddsidemargin}{0in}
    \setlength{\textwidth}{6.3in}\setlength{\topmargin}{-0.1in}}

 \def\a{\alpha}               \def\G{\Gamma}
 \def\d{\delta}      \def\D{\Delta}  \def\ep{\varepsilon} 
 \def\k{\kappa}      \def\l{\lambda}      \def\m{\mu}
 \def\f{\phi}            \def\vv{\varphi}    
 \def\j{\psi}            \def\r{\varrho}     \def\s{\sigma}  
 \def\t{\tau}

     \def\OO{{\mathcal O}}  \def\NN{\mathcal{N}}    
 \def\pa{\partial} \def\ra{\rightarrow} 
  
 \def\dd{{\rm d}}  \def\bra{\langle}   \def\ket{\rangle}

 \def\qu{\ {\buildrel {\displaystyle ?} \over =}\ }  
 
 \def\iss{\ =\ }

 \def\fract#1#2{{\textstyle{#1\over#2}}}
 \def\ffract#1#2{\raise .2 em\hbox{$\scriptstyle#1$}\kern-.3em/
                 \kern-.2em\lower .15 em \hbox{$\scriptstyle#2$}}
 
 \def\half{\fract12}  
 
 \def\part#1#2{{\partial#1\over\partial#2}}

			\def\E{\epsilon}		

\def\ds{\displaystyle}

\def\bmatrix{\begin{matrix}} \def\ematrix{\end{matrix}} \def\bpmatrix{\begin{pmatrix}}\def\epmatrix{\end{pmatrix}}
\def\bcenter{\begin{center}} \def\ecenter{\end{center}}

\def\lowerheightfig#1#2#3{\(\raise-#1\hbox{\includegraphics[height=#2]{#3}}\)}
\def\lowerwidthfig#1#2#3{\(\raise-#1\hbox{\includegraphics[width=#2]{#3}}\)}

\publishversion 

\begin{document} \begin{titlepage}

\title{\normalsize \hfill ITP-UU-12/25 \\ \hfill SPIN-12/23
\vskip 10mm \Large\bf Discreteness and Determinism in Superstrings}

\author{Gerard 't~Hooft}
\date{\normalsize Institute for Theoretical Physics \\
Utrecht University \\ and
\medskip \\ Spinoza Institute \\ Postbox 80.195 \\ 3508 TD Utrecht, the Netherlands \smallskip \\
e-mail: \tt g.thooft@uu.nl \\ internet: \tt
http://www.phys.uu.nl/\~{}thooft/}

\maketitle

\begin{quotation} \noindent {\large\bf Abstract} \medskip \\
Ideas presented in two earlier papers are applied to string theory. It had been found that a deterministic cellular automaton in one space- and one time dimension can be mapped onto a bosonic quantum field theory on a 1+1 dimensional lattice. We now also show that a cellular automaton in 1+1 dimensions that processes only ones and zeros, can be mapped onto a fermionic quantum field theory in a similar way. The natural system to apply all of this to  is superstring theory, and we find that all classical states of a classical, deterministic  string propagating in a rectangular,  \(D\) dimensional space-time lattice, with some boolean variables on it, can be mapped onto the elements of a specially chosen basis for a (quantized) \(D\) dimensional superstring. This string is moderated (``regularized")  by a 1+1 dimensional lattice on its world sheet, which may subsequently be sent to the continuum limit. The space-time lattice in target space is \emph{not} sent to the continuum, while this does not seem to reduce its physically desirable features, including Lorentz invariance. We claim that our observations add a new twist to discussions concerning the interpretation of quantum mechanics, which we call the cellular automaton (CA) interpretation. Detailed discussions of this interpretation, and in particular its relation to the Bell inequalities, are  now included.

\end{quotation}

\vfill \flushleft{July 15 and September 15, 2012}

\end{titlepage}
\eject \setcounter{page}{2}

\newsecl{Notation}{notation}

Let us briefly recapitulate our special notation: variables described by
		\be\hbox{capital Latin letters,}\qquad &N,\  P,\  Q,\  X,\  \cdots,\!\! &\hbox{ denote integer valued fields,}\qquad\quad\crl{integers}
		\hbox{lower case Latin letters,}\, \ &   p,\ q,\ \cdots, &\hbox{indicate  real numbers, and}\qquad\crl{real}
		\hbox{lower case Greek letters,}\  &  \a,\ \eta,\  \xi,\  \cdots, &\hbox{are numbers defined \textit{modulo} 1,} \eel{Greek} 
the latter being usually constrained to the interval \([\,-\half,\ \half\,] \). There will be a few exceptions, such as the lattice coordinates \(x,\,t\) on the world sheet, which are in lower case because they are not used as field variables, and the Greek letter \(\s\) and the three Pauli matrices \(\s_1,\,\s_2,\,\s_3\), which have the eigenvalues \(\pm 1\).

Frequent use will be made of the number
		\be  \E\equiv e^{2\pi}\approx\hbox{ 535{\small .49}}\,\cdots\ ,\ \hbox{ so that }\ 
		e^{2\pi i \a}\equiv\E^{i \a}\ ,\quad\hbox{and }\ \E^{iZ}=1\quad \hbox{if}\quad Z\in\mathbb Z\ . \eel{Exp}
If we have integers \(Q,\ P,\,\cdots\), we will often associate a Hilbert space of \emph{states} to these: \(|\,Q,\,P,\,\cdots\ket\). Then, there will be operators \(\eta_Q,\,\eta_P,\,\cdots\), defined by
	\be \E^{\ds iN\eta_Q}|\,Q,\,P,\,\cdots\ket=|\,Q+N,\,P,\,\cdots\ket\ ;\qquad  \E^{\ds iN\eta_P}|\,Q,\,P,\,\cdots\ket=|\,Q,\,P+N,\,\cdots\ket\ .\quad  \eel{etadef} 
These operators play the role of ``position" operators for the discrete ``momentum" variables \(Q\) and \(P\).  In general, they will have eigenvalues restricted to be in the interval \((-\half,\,\half\,]\). 

Hermitean operators will be called real or integer if all their eigenvalues are real or integer.
Because we wish to keep factors \(2\pi\) in the exponents (to be absorbed when we use \(\E\) instead of \(e\)), our commutation rules will be
	\be[q,\,p]=i/2\pi\ , \eel{xpcomm}
and the quantum evolution equation for an operator \(\OO\) is
	\be\fract\dd{\dd t}\OO=-2\pi i[\OO,H]\ , \eel{Schro}
all of which means that it is \(h\) rather than \(\hbar\) that we normalize to one. Many of the above quantities can be taken to be functions of the world sheet space-time coordinates \((x,\,t)\) or the world sheet lattice sites \((x,\,t)\).

\newsecl{Introduction}{intro}

Quantum mechanics has become a phenomenally powerful and successful doctrine for the description of all physical properties of tiny objects, whenever their sizes and mass or energy scales are near or below those of a large molecule. It appears to be universally accepted that quantum mechanics requires logical reasoning that at least deviates from standard human experience when we think of soccer balls, automobiles, stars, planets and live creatures such as cats. Attempting to ``explain" quantum logic often drives researchers into ways of reasoning that transcend ordinary logic even further. The idea that quantum phenomena might be explained in terms of totally classical, down-to-earth, underlying theories, where Hilbert spaces, tunneling, interference, and so on, play no role whatsoever, is often categorically dismissed. ``Classical thinking" belongs to the nineteenth century when people did not know any better.

There seem to be sound reasons for this attitude. Bell's inequalities\cite{Bell} and similar observations\cite{epr}\cite{CK} are applied with mathematical rigor to prove that quantum mechanics will be the backbone of all theories for sub-atomic particles of the future, unless, as some string theorists have repeatedly stressed, ``an even stranger form of logic might be needed"\cite{Gross}.

The author of this paper takes a minority's point of view\cite{GtH}\cite{Blasone}, which is that, in order to make further progress in our understanding of Nature, classical logic will have to be restored, while at the same time our mathematical skills will need to be further improved. We have reasons to believe that the mathematics of \emph{ordinary} statistics can be rephrased in a quantum mechanical language and notation; indeed this can be done in a quite natural manner, such that one can understand where quantum phenomena come from.

Our starting point is that there are underlying theories where the physical degrees of freedom are described as discrete bits of data. Some kind of time variable can be used in terms of which these data evolve. The evolution resembles a computer program, and usually we will assume that this program processes the data locally. In computer science such a system is known as a ``cellular automaton"\cite{Wolfram}.

On the one hand, we emphasize that there is no need for the reader to accept our general philosophy or interpretation. This paper was intended just to report about some simple mathematical transformations. On the other hand, however, we experienced such a fierce opposition in discussions with colleagues, that, in this new version of this paper, much more extensive discussions of our interpretation, ``hidden variables", and our axiomatic formalism  are added, see Sections~\ref{CAinterpr.sec}---\ref{axioms}. A reader who has already tears of disbelief in his/her eyes, should first consult those Sections, in particular Section~\ref{hidden}, where also Bell's theorem is adressed. For the others, let me briefly summarize.

Most, if not all, treatises on ``hidden variables" begin with formulating what is considered to be utterly reasonable assumptions as to what such variables should do: somewhere in their equations, they should, `of course', not be quantum mechanical, but they should `mimic' quantum behavior. For instance, if in genuine quantum mechanics two kinds of measurements cannot be made at the same time, the hidden variables should be able to do this anyway.

The authors then continue, to demonstrate with beautiful mathematical rigor, that such variables \emph{indeed} disagree with quantum mechanical predictions.

These assumptions do not apply to this work. The variables in this paper disagree \emph{nowhere} with quantum mechanics. They happen to be classical and quantum mechanical at the same time. There is a simple mathematical transformation from one picture to the other, and back. Therefore, if quantum mechanics forbids two measurements to be done at the same time, then this theory forbids this as well. The beautiful mathematical \emph{no go} theorems do not apply here. Their assumptions, as usually formulated on page 1, line 1, are invalid. Yet the phrase `hidden variables', to some extent, does not seem to be inappropriate here. In some extremely important ways, however, the theory adopted here differs from the usual `hidden variable'   scenario. The escape from Bell's inequalities occurs through the route of space-like vacuum correlations, and the argument is quite delicate.  A reader who is unable to accept all this, please first continue with Section~\ref{CAinterpr.sec}.

Now please allow me to explain how the ideas were conceived.

Motivated by the conviction that this is the only way to go, this work is a continuation of a new series\cite{GtH1}\cite{GtH2} where our mathematical doctrine is developed. What we wished to demonstrate is that a theory in which states evolve as in a cellular automaton can turn into a quantum field theory, by doing nothing more than a couple of mathematical transformations. While a cellular automaton undergoes its (classical) evolution processes, its observables can be cast into a Hilbert space of states, where operators can be defined just as in quantum mechanics. These operators then obey a Schr\"odinger equation. This equation itself is easy to derive and quite obvious, but the `quantum miracles' it leads to are due to extremely subtle features of the vacuum state.

So we easily arrive at the Schr\"odinger equation, which is linear, and it acts in a genuine Hilbert space. This linearity invites us to consider superpositions of states, and to pass on to a new basis of elementary states, in which a wave function evolves. What happens in terms of the original automaton is in principle very simple. The wave function that we use is a superposition of cellular automaton states, and this means that we have a probabilistic distribution there. At the level of this automaton, the \emph{phases} of the wave functions mean nothing at all, and this has no consequences either, since, in terms of the automaton states, no interference takes place. But by the time we have become accustomed to the basis elements of the quantum field theory, these seem to evolve in a much more contrived way, so that interference phenomena do seem to be important.  One of the reasons why we experience our world as if it is quantum mechanical, is that we have not (yet) been able to identify the ``ontological" basis of the cellular automaton states. 

In the terminology used in Ref.~\cite{hardy}, our theory is \emph{\(\j\)-epistemic} rather than \emph{\(\j\)-ontic}. The ontological variables, or ``hidden" variables \(\l\) discussed there, in our theory only refer to the \emph{basis elements} of a quantum theory, of only one very specially chosen basis.

The author's opinion is that the findings reported here shine a new light on the question of interpretation of quantum mechanics, but before going into any discussion about ontological or epistemic states, EPR experiments or Bell's inequalities, we must display as clearly as we can the technical calculations performed here. Extensive  discussions are added at the end. The reader should then be able to judge for him- or herself whether this affects his or her view on quantum mechanics. 

The idea is simple, and now we claim that it works. Indeed, it leads to something of a surprise. Let me disclose the surprise right away. We found a simple cellular automaton that describes strings evolving classically on a \(D\) dimensional space-time lattice. The fact that the lattice is discrete implies that, locally, bits and bytes of information are processed. Putting this string on a world sheet that is also a lattice (in one space- and one time dimension), we get an ordinary, classical field theory of integers on this lattice. The strings may also carry Boolean degrees of freedom  (fields taking the values \(\pm 1\) only) on its lattice elements. The system is integrable classically, and we find streams of left-going and right-going integers; this is the cellular automaton. It could be seen as the simplest kind of string theory one can imagine; at this level, there is no word about quantization, let alone bosons, fermions or supersymmetry.

But then, the procedure described in Ref\cite{GtH2} allows us to \emph{transform} this simple automaton into a quantum field theory of bosons and fermions. There are left-movers and right-movers, and there is a lattice cut-off. The cut-off does not affect the particle dispersion law: all modes with momentum below the Brillouin zone move exactly with the (worldsheet) speed of light. There is no direct interaction \emph{yet}. We did not (yet) consider boundary conditions, so the string has infinite length. Thus, apart from the lattice cut-off in the world sheet, this is a quantum string. After the transformation described in Ref.~\cite{GtH2}, the space-time lattice disappears and now seems to look like a continuum. Assuming the usual gauge fixing on the world sheet, we are left with \(D-2\) degrees of freedom,  the transverse coordinates.

In a similar fashion, to be described in Section~\ref{fermions}, the Boolean degrees of freedom can be transformed into fermionic fields, and thus we arrive at bosons and fermions. Indeed, if we take \(D-2\) of these fields, they will combine with the bosons to form  \(\NN=1\) super multiplets, just as in the superstring. This way, we discover that our simple discrete cellular string system turns into a superstring moving around in a space-time continuum instead of a lattice. \emph{Physically}, our discrete, deterministic lattice theory is \emph{identical} to the continuous, quantum mechanical (super)string theory; one is just a mathematical transformation of the other. The identification however, is mathematically rather contrived; we postpone further discussions  to later sections.

The original target space-time turns into a lattice with fixed lattice length, \(a_\ell=2\pi\sqrt{\a'}\), where \(\a'\) is the string slope parameter. The target space therefore cannot be sent to the continuum limit; yet it is  mathematically transformed  to become a continuous space, in fact, a superspace. 

In superstring theory, it is well-known that such a theory does not automatically reproduce Lorentz invariance in \(D\) space-time dimensions. For this, we need to have exactly 8 transverse dimensions, that is, a \(D=10\) dimensional space-time. 

As for the lattice in the world sheet, the situation is more subtle. In principle, our intention was to keep it discrete also, since the target space lattice (the lattice in space-time) induces a natural looking discrete lattice on the world sheet. Nevertheless, in a more advanced treatment of this theory, the world sheet lattice will probably have to be sent to the continuum after all. This may be needed in order to get a good, Lorentz invariant theory. To construct Lorentz boost operators, the longitudinal variables are needed, and the algebra of the Lorentz group should harmonize with the algebra of these constraints. This algebra requires world sheet reparametrization invariance, and this can be treated more easily if we first pass to the continuum limit, as in the conventional theory. Physically, going to the continuum limit on the world sheet does not modify the theory; mathematically, it is a gauge transformation, using the large amount of freedom we have in choosing the conformal world sheet coordinate transformations.

At first reading of this paper, one could decide first to read the discussions at the end, Sections~\ref{CAinterpr.sec} ---\ref{disconc}, to see where we intend to end up. Not all problems have been solved, because some of the math tends to become complicated. Boundary conditions, needed to describe open and closed strings of finite length, and GSO projections, needed to eliminate tachyon states, are hardly toughed upon. 

While reading the rest of the paper, one is advised also to inspect the two previous papers that are often referred to\cite{GtH1}\cite{GtH2}. We think the implications of this work are exciting and worth further intensive study.

\newsecl{Bosonic String Theory on the lattice}{bosstringlattice}

Consider a rectangular lattice in both space and time.  As already emphasized, we do not plan to send  the physical value \(a_\ell\) of the lattice link size  to zero at any stage of the theory. Indeed, it will be fixed by the string slope parameter \(\a'\), whose value is not yet known.  We will normalize \(a_\ell\) to one. Thus, in our space-time units, the \(D-1\) space-like lattice coordinates \(X^i\) and one time-like lattice variable \(X^0\) will all take integer values. We \emph{never} Wick rotate to Euclidean space-time.

On our discrete space-time, we describe a propagating string as follows. Besides the space-time lattice, take also a discrete, 1+1 dimensional world sheet lattice described by integer coordinates\fn{In most treatises on string theory, these are called \(\s\) and \(\t\). But for notational consistency in this paper, we prefer to use \(x\) and \(t\).}  \((x,\,t)\), and integer functions \(X^\m\) on this lattice. We could limit ourselves to the sites where \(x+t\) are even. The initial state is defined at  the world sheet time values \(t=0\) and \(t=1\) only: \(X^\m(x,0) \)and \(X^\m(x,1)\). ``String bits" \(A_L^\m\) now connect every point \((x,t)\) with \((x+1,\,t+1)\) while \(A^\m_R\) connect the points  \((x,t)\) with \((x-1,\,t+1)\) . Thus, we have
	\be X^\m(x+1,\,t+1)-X^\m(x,\,t)&=&A_L^\m(x+\half,\,t+\half)\ , \crl{leftbits}
		X^\m(x-1,\,t+1)-X^\m(x,\,t)&=&A_R^\m(x-\half,\,t+\half)\ , \eel{rightbits}
Thus, apart from the center of mass, the entire set of initial values of the string is determined by the set of numbers \(A_L^\m(x,\,\half)\) and \(A^\m_R(x,\,\half)\) at all \(x\).

These vectors will be chosen to be composed of integers, and strictly light like (later, we may replace this constraint by a different one):
	\be (A_L^\m(x,\,\half))^2 =0\ ,&&(A_R^\m(x,\,\half))^2 =0\ ,\crl{Aconstraint}
	 A_L^0(x,\half)\ge 0\ ,&& A_R^0(x,\half)\ge 0\ .\eel{Azerobound}
This ensures that, at \(t=0\) as well as at \(t=1\), the initial data \(X^\m(x,t)\) are not timelike separated from both neighbors \(X^\m(x\pm 2,\,t)\).

We now postulate the equations of motion:
	\be X^\m(x,t)=X^\m(x-1,\,t-1)+X^\m(x+1,\,t-1)-X^\m(x,\,t-2)\ . \eel{discreom}
One readily derives that \(A^\m_L\) are left-movers and \(A^\m_R\) are right-movers:
	\be A^\m_L(x,\,t)=A^\m_L(x+t)\ ,\qquad A^\m_R(x,\,t)=A^\m_R(x-t)\ , \eel{leftrightmovers}
and therefore, the constraints \eqn{Aconstraint} and \eqn{Azerobound} hold at all times \(t\).

The careful reader may recognize these lattice equations as a discrete version of the bosonic relativistic string equations\cite{GSW}\cite{Pol}, 
	\be (\pa_x^2-\pa_t^2)X^\m=0\ ;\qquad (\pa_x X_L^\m(x+t))^2=0\ ,\quad(\pa_xX_R^\m(x-t))^2=0\ , \eel{conteom}
which, in the continuum limit \(a_\ell\ra 0\), would yield the conventional, classical theory. In this paper, however, we will show that \(a_\ell\) is linked to the square root of the slope parameter \(\a'\) of the string, which we will \emph{not} send to zero. Furthermore, in spite of the appearances from Eq.~\eqn{discreom}, our theory will show to be a quantum theory, not a classical one (Note, that Eqs.~\eqn{conteom} hold for the quantum theory as well). The lattice artifacts will not be observable, even at the Planck scale.

A very important property of the classical equations \eqn{leftbits}---\eqn{leftrightmovers} is that they are invariant under the discrete Lorentz group \(O(D-1,1,\,\mathbb Z)\), that is, the Lorentz group restricted to integers in its operator matrices. At first sight, these bosonic strings seem to be non-interacting, but one interaction mode, also invariant under  \(O(D-1,1,\,\mathbb Z)\), can be introduced:  an exchange operation: if two strings hit the same space-time point \(X^\m\), their arms are exchanged. This may generate closed strings. The author has long been searching for classical lattice theories, with interactions, invariant under  \(O(D-1,1,\,\mathbb Z)\). Here, finally, we have one. Note that, our theory requires also the interactions to be classical, and this requires the exchange interaction to be unambiguous; consequently, the bosonic string described here has to be an \emph{oriented} string. And, its string interaction constant, \(g_s\), \emph{cannot} be tuned to any value; it is basically equal to one.

\newsecl{Bosonic string theory in continuous space-time}{bosstrcont}

As is standard in string theories, we see that \(A^0_{L,R}\) fixes the physical time coordinates, and Eqs.~\eqn{Aconstraint} remove  one more spacial degree of freedom, so that, effectively, there are only \(D-2\) free parameters left at every lattice site \((x,t)\). These obey the linear equations \eqn{discreom}. While the \(X^a(x,t),\ a=1,\cdots, D-2,\) are independent integers, we have exactly the discrete cellular automaton described in our previous paper \cite{GtH2}, in \((D-2)\)-fold. It is described there how this system is \emph{equivalent} to a quantum theory of \(D-2\)  quantum fields \(q^a(x,t)\), and associated field momentum variables \(p^a(x,t)\). Both \(q^a\) and \(p^a\) take \emph{real} values, while still living on a world sheet lattice. 
They obey the quantization rules:
	\be[q^a(x,t),\,p^b(x',t)]= \fract{i}{2\pi}\d^{ab}\d_{x,x'}\ . \eel{qpcomm}
We can summarize the result of Ref.~\cite{GtH2} as follows. The equivalence transformation starts with the left- and right-movers,  \(A^a_{L,R}(x)\). The commutation rules for the quantum operators \(a^a_{L,R}(x)\), associated to the classical, discrete functions \(A^a_{L,R}(x)\), must be\fn{with apologies for using the letter \(a\) both for counting the \(D-2\) transverse degrees of freedom and for the real valued left and right moving fields. Confusion is hardly possible.}
	  \be[a_L^a,a_R^b]=0\ ;&& [a_L^a(x),\,a_L^b(y)]=\pm\fract i{2\pi}\d^{ab}\ \hbox{ if }\ y=x\pm 1\ ;\quad\hbox{else }\ 0\ ; \crl{aLaLcomm}
				&& [a_R^a(x),\,a_R^b(y)]=\mp\fract i{2\pi}\d^{ab}\ \hbox{ if }\ y=x\pm 1\ ;\quad\hbox{else }\ 0\ . \eel{aRaRcomm}

 Let us concentrate on the left-movers, and drop the subscript \(L\) and the superscript \(a\).  They only live on the odd sites \(x\). Since the variables \(A(x)\) are integers, we regard them as discrete ``momentum" operators associated to  ``position"  operators \(\eta_A(x)\). Just as in Section~\ref{notation}, we demand the operators \(\eta_A(x)\) to be periodic with period 1, so that these variables at all points \(x\)  together generate a torus with \(\ell\) periodic dimensions, if \(\ell\) is the total length of a string. As explained in Ref.~\cite{GtH2}.
We could define the continuous quantum operators \(a(x)\) as
	\be a(x) \qu A(x)-\eta_A(x-1)\ , \eel{aAidentificationattempt}
which would obey the correct commutation rules, except where \(\eta_A(x)=\pm\half\), because the jump from \(+\half\) to \(-\half\) is not accounted for. A better definition is 
	\be a(x)\iss -\fract i{2\pi}\,{\pa\over\pa\eta_A(x)}\ +\ {\pa\over\pa\eta_A(x)}\,\vv(\{\eta_A(x)\})\ -\,\eta_A(x-1)\ , \eel{aAidentification} 
where the phase function \(\vv(\{\eta(x)\})\) is a carefully chosen functional\cite{GtH2}, in such a way that \(a(x)\) depends continuously on \(\eta(x)\) and \(\eta(x-1)\) at the cross-over from \(\half\) to \(-\half\), although this does leave an inevitable singularity when \emph{both} \(\eta(x)\) and \(\eta(x-1)\) are \(\pm\half\).

Thus, singularities are generated at the boundary points of two successive \(\eta\)'s, the points where \(\eta_A(x)=\pm\half,\ \eta_A(x-1)=\pm\half\). At these points, wave functions must vanish in order to keep the commutation rules \eqn{qpcomm} valid\fn{In Ref.~\cite{GtH2}, the suspicion was expressed that this constraint could be removed by introducing fermionic degrees of freedom. We refrain from continuing along that route, however, since, although we did find that fermions may remove this constraint, they in return then deliver even more troublesome anomalies in the fermionic (anti)commutation rules. Fermions will be introduced in the next section.}.

Let us put the indices \(L,R\), and \(a\) back. Then, the quantum operators \(q^a(x,t)\) and \(p^a(x,t)\) are defined by the equations
	\be a_L^a(x+t)&=&   p^a  (x,t)+ q(x,t)- q^a(x-1,t)\ ;\crl{aL}
	     a_R^a(x-t)&=&   p^a  (x,t)+ q(x,t)- q^a(x+1,t)\ , \eel{aR} 
which can easily be solved by Fourier transformation in the \(x\) coordinate.

We note that the quantum variables \(q^a(x,t)\) are the transverse string variables usually called \(X^a(\s,\t)\) in string theory, \(p^a(x,t)=\pa_\t X^a(\s,\t)\), and  they obey the usual commutation rules. The only thing unusual to them is that the world sheet is still a lattice\cite{Suss}. However, as was also emphasized in Ref.~\cite{GtH2}, the hamiltonian at world sheet momentum values \(\k\) with \(|\k|<\half\), is \emph{identical} to the continuum hamiltonian, so the only way in which the world sheet lattice manifests itself here is in the presence of a Brillouin limit to the momentum, while at values of \(\k\) within the Brillouin zone, the continuum theory is exactly reproduced.

How much does the theory obtained here actually differ from the strictly continuous string theory? As long as we concentrate on the transverse modes only, there is reason to suspect that there is no difference at all. To describe highly excited strings, the high \(\k\) values may not be needed at all; we just go to longer strings, an observation already made by Klebanov and Susskind in Ref.~\cite{Suss}. To be precise, if we limit ourselves momentarily to the description of the bulk string, we can always use local conformal
transformations on the world sheet to ensure that the values of the left- and right movers \(a^\m_{L,R}\) stay within bounds and are slowly varying.  

We do note that, the constraints~\eqn{Aconstraint} and \eqn{Azerobound} do not quite coincide with the constraints usually applied in the quantum theory, since the latter are defined in terms of the complete operators \(a(x)\), not only their integral parts. The consequences of this for Lorentz invariance and the usual string anomalies are yet to be investigated (see also Subsections~\ref{pplus} and \ref{strconstr}).
	
\newsecl{Fermions}{fermions}

Now consider the same world sheet lattice \((x,t)\), but now with Boolean degrees of freedom on them: \(\s(x,t)=\pm 1\). Let the classical equation of motion be similar to \eqn{discreom}:
	\be\s(x,t)\iss\s(x-1,\,t-1)\ \s(x+1,\,t-1)\ \s(x,\,t-2)\ . \eel{booleom}
Again, we have left movers and right movers:
	\be\s(x+1,\,t+1)\ \s(x,t)&=&\s_L(x+t+1)\ ,\crl{leftsigma} \s(x-1,\,t+1)\ \s(x,t)&=&\s_R(x-t-1)\ .\eel{rightsigma}
Identifying these operators with the Pauli \(\s^3\) operators \(\bigg(\matrix{1&0\cr 0&-1}\bigg)\), we also introduce the operators
	\be\s^1=\bigg(\matrix{0&1\cr 1&0}\bigg)\ ,\ \hbox{and }\  \s^2=\bigg(\matrix{0&-i\cr i&0}\bigg)\ , \eel{pauli}
so that we have \(\s^i_{L,R}(x)\ , \ i=1,2,3\), obeying
	\be[\s^i_L(x),\,\s^j_L(x')]=2i\d_{x,x'}\,\ep_{ijk}\,\s^k_L(x)\ ,\qquad [\s^i_L(x),\,\s^j_R(x')]=0\ , &&\crl{paulicomm}
	\hbox{and }\quad\{\s^i(x),\,\s^j(x)\}\equiv\half\bigg(\s^i(x)\,\s^j(x)+\s^j(x)\,\s^i(x)\bigg)   =\d^{ij}\ .&&\eel{paulianticomm}
In 1+1 dimensions, we wish to use fermionic fields \(\j_A(x),\ A=1,2\), obeying
	\be \j_A^\dag(x)=\j_A(x)\ ,\qquad    \{\j_A(x),\j_B(x')\}=\d_{x,x'}\d_{AB}\ . \eel{psianticomm}
They can be obtained from the operators \(\s^i_{L,R}(x)\) for instance as follows:
	\be\j_1(x)=\j_L(x)=\s^1_L(x)\prod_{y<x}\s^3_L(y)\ ,\quad\j_2(x)=\j_R(x)=\s^1_R(x)\prod_{y<x}\s^3_R(y)\,\prod_z\s^3_L(z)\ . \eel{JW}
This is the so-called Jordan-Wigner transformation\cite{JW}.	It is non-local only if we take products of an odd number of \(\j_L\) or \(\j_R\) operators. In practice, this is never needed.

Consider now the Fourier transforms of these fermionic fields:
	\be\hat\j_A(\k)=\sum_x\j_A(x)\,\E^{-i\k x}\ ,&&\j_A(x)=\int_{-\half}^\half\dd\k\, \hat\j_A(\k)\E^{i\k x}\ , \crl{fourierpsi}
	\hat\j^\dag_A(\k)=\hat\j_A(-\k)\ ,&&\{\hat\j_A(\k),\,\hat\j_B(\k')\}=\d(\k+\k')\ ,\eel{fourieranticomm}
and introduce the lattice hamiltonian operator\fn{The factor \(\half\) arises from the normalization chosen in Eq,~\eqn{paulianticomm}.} 
	\be H=\half\int_{0}^\half\dd\k\,\k\,\hat\j_A(-\k)(-\s^3_{AB})\hat\j_B(\k)\ ; \eel{fermiH}
this gives
	\be[H,\hat\j_A(\k)]=-\s^3_{AB}\,\k\,\hat\j_B(\k)	\ . \eel{Hfermicomm}
According to the Schr\"odinger equation \eqn{Schro}, the time dependence is then
	\be\hat\j(\k,t)=	e^{2\pi i\s^3\k t}\hat\j(\k,0)\ , \eel{fouriertimedep}
so that 
	\be \j_1(x,t)&=&\E^{i\k(x+t)}\hat\j_1(\k)\iss\j_L(x+t)\, ,\crl{leftpsi}
	 \j_2(x,t)&=&\E^{i\k(x-t)}\hat\j_2(\k)\iss\j_R(x-t)\, .\eel{rightpsi}

Thus we see that the quantum hamiltonian \eqn{fermiH} produces the correct time dependence of all fermionic operators of the automaton. We now claim that it also propagates all individual \emph{states} correctly. The argument goes as follows.

Consider first the state with all ``spins" up: \(\s^3_L(x,0)=\s^3_R(x,0)=+1\). One quickly finds that, for this state, the hamiltonian vanishes. This is because, for this state, it can only use operators of the form \(\s_A^1(x)\,\s_A^1(x+N)\), but for both \(A=1\) and \(A=2\), the hamiltonian is antisymmetric under left-right interchange, so that the coefficient for all these terms must cancel out.

Next, we find that, acting on this state with an arbitrary number of \(\j\) operators, gives all other states with left- and right movers. These now propagate the same way both in the classical and in the quantum mechanical description.
	
It is a special feature of the 1+1 dimensional case that a massless fermionic lattice field theory maps onto a simple cellular automaton. In higher dimensions, this is much more difficult to achieve, because we cannot make use of left- and right movers. In Ref.~\cite{GtH2}, we found the same result for the bosonic case.

\newsecl{The superstring and its constraints}{sustr}

The locally independent degrees of freedom of the superstring are the bosonic coordinates \(q^a(x,t)\)  and the fermionic fields \(\j^a_A(x,t)\), with \(a=1,\cdots,D-2\). We have seen that these map onto the discrete integers \(X^a(x,t)\) and the Boolean variables \(\s^a(x,t)\). It is important to note, at this stage, that also the equations of motion exactly correspond to those of the superstring: all excitations move with the local speed of light to the left or to the right along the string world sheet.

However, the superstring also has two longitudinal coordinates \(q^{D-1}\) and \(q^0\), as well as two longitudinal fermionic fields, \(\j_A^{D-1}\) and \(\j_A^0\). These are not independent of the other degrees of freedom, but determined by them via two kinds of equations: \emph{gauge fixing} equations and \emph{constraints}. Even though the original string theory had a continuous world sheet while our world sheet is a lattice, we can make a rigorous identification: the Fourier modes of the left- and right movers, \(a^\m_{L,R}\) and \(\j^\m_A\), must correspond exactly with the Fourier modes of the left- and right-movers of the continuum theory, as long as \(|\k|<\half\). Higher Fourier modes of the continuum theory are set to zero. If we allow the length of the string in its world sheet to be arbitrary, one can still obtain all string excited modes, so that this Brillouin limitation is a minor one, in practice.

For the continuum theory, we now write the world sheet gauge fixing conditions as
	\be a_L^+=1\ ,\qquad a_R^+=1\ ,\qquad\j_1^+=\j_2^+=0\ ,\eel{worldgauge}
where \(a^\pm=a^0\pm a^{D-1}\), \(\j^\pm=\j^0\pm\j^{D-1}\), and the superstring constraints are:
	\be (a_L^\m)^2+\half\j_1^\m\,\hat\k\,\j_1^\m \iss 0\ ,&&\   (a_R^\m)^2+\half\j_2^\m\,\hat\k\,\j_2^\m \iss 0\ ; \crl{boseconstr}
		a^\m_L\,\j^\m_1\iss 0\ ,&&\  a^\m_R\,\j_2^\m\iss 0\ , \eel{fermiconstr}
where \(\hat\k\) is the operator that multiplies the Fourier components \(\hat\j(\k)\) with \(\k\). The latter two equations now fix the Lorentz light cone minus components of these variables, since the bosonic constraint \eqn{boseconstr} gives
 	\be a^-_L  = \sum_{a=1}^{D-2}((a^a_L)^2+\half\j_1^a\,\hat\k\,\j_1^a)\ , \eel{bosconstr}
while the fermionic constraint \eqn{fermiconstr} gives
	\be	\j^-_1&=&2\sum_{a=1}^{D-2}\j_1^a\,a_L^a\ ,\eel{fermconstr}
and the same with \(a_R\) and \(\j_2\).

The way it works in the superstring is that these constraints, as well as the longitudinal degrees of freedom that are determined by them, could be completely ignored as long as we are only interested in the string equations of motion. Whether this includes the superstring interactions requires further research but in principle this should be straightforward. The importance of the constraints lies entirely in the fact that they are needed to investigate Lorentz invariance. The generators of the Lorentz transformations are determined by all \(D\) components of the dynamic variables.

Therefore, in our mapping between discrete theories and the superstring, we may also ignore the longitudinal degrees of freedom and the constraint equations obeyed by them, except when we wish to study Lorentz invariance.

In our lattice theory, we originally expected invariance under \emph{discrete} Lorentz transformations \(O(D-1,1,\,\mathbb Z)\) by adding discrete longitudinal components to the left- and right movers \(A^a_{L,R}(x,t)\). However, what is really needed is only the gauge fixing and the constraints for the continuum case, because that is the system for which Lorentz invariance is required. The quantum constraint equations should  obey the same algebra as in the superstring with a continuous world sheet. Details of this problem will be left for further investigation.

Our conclusion of this section is that the theory has to be viewed as follows:
	\bi{*} We start with a classical (i.e. non quantum mechanical) dynamical theory that describes \(D-2\) degrees of freedom \(A^{a\,L,R}\ (a=1,\cdots,D-2)\) as integer-valued left-moving and right-moving fields on a 1+1 dimensional world sheet. These degrees of freedom act as discrete coordinates on a transverse space-like lattice. The lattice length is fixed: \(a_\ell=2\pi\sqrt{\a'}\). The theory is invariant only under discrete \(SO(D-2,\mathbb Z)\) rotations and discrete translations.
	\itm{*} By adding the canonical associated operators \(\eta^{L,R}_A\) we can use these to fill the spaces between the lattice sites so that a space-like \(D-2\) dimensional continuum is obtained. Now, however, the fields do not commute, so we obtain a quantum version of this system: a quantized string theory. The invariance group turns into that of the transverse continuous rotations and translations.
	\itm{*} A similar procedure starting with Boolean degrees of freedom gives us fermonic fields. Thus, we find our mapping connecting a deterministic automaton processing integers and Boolean variables to a (quantized) superstring theory.
	\itm{*} Only in the quantum theory, we can now follow standard procedures to use gauge conditions~\eqn{worldgauge} and constraints \eqn{boseconstr} and \eqn{fermiconstr} which define for us the longitudinal space-time coordinates.
	\itm{*} With these longitudinal coordinates added, we use standard arguments to prove invariance under the full Lorentz group \(O(D-1,1,\mathbb R)\), but only if \(D=10\). Clearly, only the quantum system possesses this manifest Lorentz invariance. In the classical automaton, this invariance is not manifest at all.
	\ei

\newsecl{The CA interpretation of quantum mechanics}{CAinterpr.sec}

\subsection{Short History}\lab{history} Intuitively simple and natural interpretations of quantum mechanics could be extremely helpful for making further progress  in understanding quantum gravity, string theory and quantum cosmology. The ``many world" interpretation, or the Bohm - deBroglie ``pilot waves", and even the idea that the quantum world must be non-local, are difficult to incorporate in models of the universe. Of course, it was attempted to adopt these view points in numerous approaches, but the resulting picture is often opaque and not in line with intuition. By itself, such impressions do not disqualify these interpretations, which indeed have become fairly standard, but it seems very much worth-while to continue searching for something better.

We should plainly ask \emph{what is really going on} in a world described by quantum mechanics, and in order to try to get some ideas,  this author constructed some models with various degrees of sophistication\cite{GtH}. These models are of course "wrong" in the sense that they do not describe the real world. They do not generate the Standard Model, but one can imagine starting from such simple models and adding more and more complicated details to make them look more realistic, in various stages.

Of course, the difficulties that arise when one tries to underpin quantum mechanics with determinism, are well-known. Simple probabilistic theories fail in an essential way. One or several of the usual assumptions made in deterministic theories will probably have to be abandoned. Of course, we are fully aware of that. On the other hand, our world seems to be extremely logical and natural.

Therefore, we decided to start an investigation at the other end. Make assumptions that later surely will have to be amended; make some simple models, compare these with what we know about the real world, and then modify the assumptions any way we like.

The \emph{no-go} theorems tell us that a simple cellular automaton model is not likely to work. One way to try to ``amend" them, was to introduce information loss. At first sight this would carry us even further away from quantum mechanics, but if one looks a little more closely, one finds that one still can introduce a Hilbert space, although it becomes much smaller and it may become holographic, but this is something we may actually want. If one then realizes that information loss makes any mapping from a deterministic model to quantum mechanical states fundamentally non-local - while the physics itself stays local - then maybe the idea becomes more attractive.

Now the problem with this is that again one makes too big assumptions, and the math is quite complicated and unattractive. So, reluctantly, we returned to a reversible, local, deterministic automaton and asked: to what extent does this resemble quantum mechanics, and where does it go wrong? The idea we had in mind was that we will alter the assumptions, maybe add information loss, put in an expanding universe, but all that comes later; first let us try to establish what goes wrong.

And here is the surprise: in a sense, nothing goes wrong. All one has to assume is that quantum states can be used, even if the evolution laws themselves are deterministic. Thus, the probability distributions are given by quantum amplitudes. The point is that, when describing a mapping between the deterministic system and the quantum system, one has a lot of freedom. If one looks at any one periodic mode of a deterministic system, one can define a common contribution to the energy for all states in this mode, and this observation introduces a large number of arbitrary conserved constants, so we are given much freedom.

Using this freedom, we end up with quite a few models that just look interesting. Starting with deterministic systems, we end up with quantum systems. These are real quantum systems, not any kind of ugly concoctions. On the other hand, they are still a far cry from the Standard Model, or even anything else that shows decent, interacting particles.

The exception now is string theory. This is puzzling. Is the model we constructed in this paper a counter example, showing that, what everyone tells us about fundamental quantum mechanics being incompatible with determinism, is simply wrong? This, we do not believe; we still believe that, somewhere, we will have to modify our assumptions, but maybe the usual assumptions made in the \emph{no-go} theorems, will have to be looked at as well.

It does seem that people are too quick in rejecting ``superdeterminism", and maybe also in denouncing ``conspiracy". Superdeterminism simply implies that one cannot ``change one's mind" (about which component of a spin to measure), by ``free will", without also having a modification of the deterministic modes of our world in the distant past. This is obviously true in a deterministic world, but we do observe that the free will postulate can be replaced by another one: the \emph{freedom to choose the initial state}. This condition is more precise, and we will discover that this freedom actually has to be constrained, in order to bypass Bell's inequality.

One has to keep in mind that what looks like `conspiracy' to one observer, might actually be a quite natural phenomenon to another. We will find that this consideration has to be taken seriously.
We took the attitude that questions of this sort can also be settled if we can construct explicit mathematical models that can account for the complex structure of our world. If the usual \emph{no-go} theorems are insurmountable, we should automatically be confronted with failures of our mathematical models. Curiously, this does not seem to be happening. We are zooming in on models that become more and more realistic. 

Let us now imagine that a really interesting quantum field theory of the world can be found that allows a mapping onto a cellular automaton (CA). Then, we would have new support for the CA interpretation of quantum mechanics. It could still be that the real world would require further modifications. We will be ready for that, but first allow us to go ahead with this interpretation.  As stated, it was greeted with considerable skepticism. Well, these skeptics would not have expected the mathematical model described in this paper, so perhaps they will react the way they should: here we have a paradox, an interesting one, and let us see what we can conclude from it. 

\subsection{The Cellular Automaton (CA) }\lab{CAinterpr}

The CA interpretation is embarrassingly simple. An automaton starting at time \(t_1\) in a state  \(\vec Q_1\), evolves and ends up in a state  \(\vec Q_2\) at \(t_2\). The evolution is deterministic. The simplest possibility would be that we have cells in 3-space, numbered by an index \(i\), containing data in the form of numbers \(Q^i\). At the beat of an external clock, the number in each cell \(i\) is replaced, according to an algorithm that would typically look at the number \(Q^i\) itself and the numbers \(Q^j\) in neighboring cells \(j\) only. In this case we would talk of a \emph{local} CA. To what extent a CA can be limited to be local remains to be seen. Such a CA often functions as a universal computer, but we also consider simpler test examples, which are easier to subject to simple mathematical analysis.

Our cellular automaton itself  \emph{never} ends up in any `quantum superposition'. Thus, its evolution is classical. It is \emph{not} obvious whether the evolution law is time reversible, but in our simplest models, it is. Note, that up to this point, we are not dealing with probabilities at all. All one has to do is compute \(\vec Q_2\), given \(\vec Q_1\).

However, to do the calculation correctly may be very difficult, if not impossible. If all this happens at the Planck scale, there are far too many cells, evolving far too rapidly. Now enters quantum mechanics. According to the CA interpretation, quantum mechanics is just a handy tool. We define a quantum Hilbert space, spanned by the states \(|\vec Q(t)\ket\), so that these states form the basis. In this Hilbert space, the evolution law can be written in operator form,
	\be \bra \vec Q_2|U(t_1,t_2)|\vec Q_1\ket=1\ \hbox{ if \ \(\vec Q_1\) evolves into \(\vec Q_2\) }\ ;\quad 0\  \ \hbox{ otherwise.}\eel{permevolve}
Causality requires that \(U(t_1,\,t_3)=U(t_2,\,t_3)\,U(t_1,\,t_2)\).
If the evolution itself is time-independent, we can write down an operator \(H\) such that
	\be U(t_1,\,t_2)=e^{-iH(t_2-t_1)}\ . \eel{hamevolve}
Such a hamiltonian is often easy to construct, and it even allows us to impose further constraints such as positivity and some version of locality. To calculate the evolution operator for large time intervals, we now make as many unitary transformations to different basis sets as we please. Irrespective of the basis, we have \((\dd/\dd t)|\j\ket=-iH|\j\ket\). Therefore, we may choose \emph{any} basis, to solve this equation. Then, transform back to the original basis. If the calculation is done correctly, this gives a state that coincides with a basis element  \(|\vec Q_2\ket\) at time \(t_2\); \emph{the outcome may never be a superposition}, such as \(\l|\vec Q_1(t_2)+\m|\vec Q_2(t_2)\ket\).

We also observe that we could have started with a totally arbitrary wave function \(|\j(t_1)\ket\), such that the probabilities for the initial states are defined by 
	\be \r(\vec Q,t_1)\equiv|\bra\vec Q|\j(t_1)\ket|^2\ .\eel{CABorn}
 The same Schr\"odinger equation as we have used before then gives us the wave function at time \(t_2\), such that the probabilities at \(t_2\) are indeed \(\r(\vec Q,t_2)=|\bra\vec Q|\j(t_2)\ket|^2\). This is exactly correct just because the CA does not produce superpositions itself.

During the calculation, we may have used non-commuting operators, and we may have noticed that the state we are interested in was an eigen state. It can never have been in eigen states of two non-commuting operators at the same time. 
And there is one restriction: there is no ontological interpretation of amplitudes
	\be\bra\j_1(t_1)|\j_2(t_2)\ket \eel{amplit}
 if neither \(|\j_1\ket\) nor \(\j_2\ket\) are basis elements. This restriction is important to remember.
 
 An other very important point to remember is that Eq.~\eqn{CABorn} \emph{cannot be inverted}: Any given probability distribution \(\r(\vec Q)\) corresponds to a very large class of different wave functions \(\j(\vec Q)\), since the phases may be chosen at will. So, given the probabilities, we may need to choose different wave functions under different circumstances, a feature often ignored in hidden variable scenarios.

Our system is quantum mechanical in the sense that we use Schr\"odinger equations to solve its dynamics. It is classical in the sense that CA states evolve into CA states. We can map all its dynamical properties from the quantum picture to the classical picture and back. The CA states are the ontological states. The wave functions 
\(|\j_i(t)\ket=|\vec Q_i(t)\ket \) are ontological (apart from their phases), and the complete set of ontological wave functions form an orthonormal basis.

Two more remarks: 1) The hamiltonian defined in Eq.~\eqn{hamevolve} is far from unambiguous (If time is limited to integer multiples of a unit \(\D t\), then we can add any multiple of \(2\pi/{\D t}\) to any of the eigen values of \(H\) without modifying the evolution, and this is not the only ambiguity). This means that one CA can be mapped onto many different quantum models. These however, are not physically different; they are related by unitary transformations. \\
2) Because of this ambiguity, and because there are very many different CAs, it may well be that CA models are \emph{dense} in the set of all quantum models. It is not true that every quantum model can be mapped onto a CA, but it is quite possible that many interesting quantum models can be mapped onto CAs with some margin of accuracy. In practice, however, one would like to reflect locality properties of the quantum system into the CA, and such a demand can make the actual construction far from easy.

\subsection{Measurements}\lab{measure}
We had made an important restriction in that the amplitudes \eqn{amplit} cannot be measured if both \(|\j_1\ket\) and \(\j_2\ket\) are not elements of the CA basis. Which measurements are possible? Here, we are going to make one more assumption. It is safe to assume that all measurements end up being macroscopic. Only those phenomena that can affect the position of a pointer on a device, and eventually the position of any macroscopic body such as a person or a planet (or a cat) are measurable. 

We make the essential assumption that, statistically, a macroscopic object influences the distribution of ontological values \(|\vec Q(t)\ket\) of the automaton. This means that, measuring \(|\vec Q(t)\ket\) sufficiently carefully gives us all classical data we want, and with that, the outcome of all measurements. 

We emphasize now that this should also include all quantum measurements, since any measurement must end up being macroscopic. Imagine now that we have employed some fancy quantum Hilbert space to calculate the evolution transition amplitudes \eqn{permevolve}. If we succeed in designing an apparatus that turns a quantum phenomenon into a macroscopic phenomenon, then this is classically measurable, and only the transition amplitudes \eqn{permevolve} suffice to detect the outcome; the amplitudes \eqn{amplit} are not needed.

Thus, imagine that we have constructed a CA model that can be mapped onto the Standard Model in accordance with our prescriptions, then we can perform all measurements that we are accustomed to, as soon as  macroscopic detection devices such as particle detectors can be constructed out of the atoms of this model, since its pointers are classical. A classical object is then defined by requiring that it affects the statistical distributions of the CA to such an extent that it can be distinguished from all other objects. 

Opponents of our theory now bring forward that this means that not all expectation values of all operators can be measured. At first sight this seems to be so, but remember that we do have a system that obeys exactly the Schr\"odinger equations of the full, interacting, quantum field theory. This implies that, at various occasions, systems may seem to be in eigenstates of operators that do not commute with CA observables, while in reality they do lead to classical effects on pointers of measuring devices. These effects do commute with the CA observables, and that is what matters, in the end.

This viewpoint is quite natural and conventional in quantum field theories. Here, one can derive that, given \emph{any} type of field \(\f(x)\), where \(x\) is a space-time point in Minkowski space, then the set of vacuum expectation values
	\be \G(x_1,\cdots,\,x_n)\equiv\bra\,\emptyset\,|\f(x_1)\,\cdots\ \f(x_n)|\,\emptyset\,\ket\ , \eel{npoint}
suffices to determine all elements of the entire scattering matrix, including expectation values involving any of the other fields. 
This is because, in moment space, the connected parts of these amplitudes display poles due to all the other field variables, which one can single out to obtain these other amplitudes.
In our theory, we simply take one of the CA fields to play the role of \(\f(x)\).

Thus, the bottom line is that, since all measurements are classical, all measurable operators end up to commute with the CA operators, and hence they all commute. What we are saying here is that quantum superpositions will never stay visible as such after a measurement has been done.

This also explains how an ontological wave function ``collapses": since classical observables all commute with the CA observables, the ontological wave functions are always diagonal in the classical observables and never in a superposition: Schr\"odinger's cat is alive \emph{or} dead!\cite{GtHcollapse}

In practice, a measurement can never do more than produce an estimate of the probability distribution \(\r(\vec Q)\). The manifold defined by these probabilities has a dimensionality that is only half of that of the Hilbert space of wave functions. This means that many different wave functions produce exactly the same probability distribution. We not only cannot  determine the wave function after an experiment has been performed, but we also cannot prepare a state with a given wave function. The best we can do to obtain information on the wave function that we wish to detect or produce, is to mimic the construction of a measuring device, such as in a Stern-Gerlach experiment, or by constructing macroscopic polarization filters. According to the quantum field theoretical argument given above, such devices can be imagined if we have sufficiently non-trivial interactions in our model. 

One does not have to believe that the CA interpretation is right. Or perhaps it is nearly right:  it may well be that one ingredient, not used in the present work, should actually be included to obtain a workable theory: classical information loss\cite{mathbasis}. It could also be that the expanding universe plays an essential role here, since without expansion, it would have been difficult to imagine how the necessary cooling mechanism would have taken place, while this would be crucial to understand why we are surrounded by a near vacuum. The vacuum will play an important role, as we shall see.

Taking all counter arguments in consideration, it may seem difficult to defend the CA interpretation, without any explicit model that allows us to test the various assertions. This is why we find the construction of the model in this paper important, even if some ingredients may still be missing, such as a detailed mathematical treatment of the superstring constraints and the superstring interactions. But now that we seem to be homing in to exactly the CA model we were searching for, it is of importance to describe more precisely what the CA interpretation says. We begin with what it does not say.

\newsecl{Hidden variables}{hidden}
	The usual \emph{no-go} theorems concerning hidden variables are based upon assumptions concerning the theory they plan to disprove. It is assumed that the purported hidden variables can do the impossible: either provide for the outcome of ``counterfactual" measurements, or compare the outcome of experiments made under exactly the same initial conditions, except for one change made in the vicinity of one point in space, but later in time.
	
	Counterfactual measurements are measurements that cannot be made, so a good hidden variable theory should not be required to give the outcome of such a measurement. Secondly, an exact repetition of an experiment with exactly the same initial conditions (including the values of all ``hidden variables") is impossible, if one wants to change the situation near one point later in time.  A deterministic theory does not \emph{have to} define what the outcome of such measurements should be, as they are ill-defined. Instead, the theory should tell us what \emph{does} go on in a Bell experiment. 
	
	The cellular automaton (CA) interpretation of quantum mechanics, the backbone philosophy of this paper, pretends to do exactly that.

\subsection{Vacuum fluctuations}\lab{vacuum.sec}

First, we have to handle a complication, and indeed an essential one. It  is that, in the world we are familiar with, all objects are surrounded by regions that we call \emph{vacuum}. The vacuum state is a unique eigen state of the hamiltonian with eigen value zero. All other states usually considered are very near to that zero eigen state, if just a few light particles are around. This means that all states of interest will have to be handled as superpositions of a large number of --- perhaps a sizeable fraction of all --- CA states. The vacuum can only be regarded as ontological states rapidly evolving into one another. Requiring the hamiltonian to be in its lowest eigen states means that, here, the relative phases are of importance (they are dictated by our choice of hamiltonian), and the word `vacuum fluctuations' aptly describes this situation. 

In spite of this chaotic picture of what the vacuum looks like in terms of ontological states, it is in reality a very special state. There will be quite complex, `entangled', correlations between the many possible values of the CA's memory cells, and it is important to realize that these correlations will be non-trivial at spacelike distances. As soon as the correlations differ from the vacuum ones in a statistically significant manner, we have a state with some classical object around. This is the way in which we will characterize physical observations. If we have significant deviations from the vacuum fluctuations, we have a state \(|\j\ket\) that will be orthogonal to the vacuum, and it will be recognizible by operators that are \emph{diagonal} in the CA basis. This is a crucial observation to make the idea that such operators suffice to do just any observation or measurement, more acceptable.

How it can be that such a vacuum state came about in our world is a question that we will not answer completely. One may imagine that the universe started out with a Big Bang at infinite temperature. It expanded, and with that, cooled. How the universe came to be as it is, dominated by regions of vacuum, is then the subject of cosmology.

\subsection{What happens in a Bell experiment?}\lab{Bell.sec}
In Bell's paper\cite{Bell}, the starting assumption is that a hidden variable, \(\l\), provides for quantities \(A(\vec a,\,\l)\) and \(B(\vec b,\,\l)\), such that the expectation value for a combined measurement is
	\be P(\vec a,\,\vec b\,)=\int\dd\l\,\r(\l)\,A(\vec a,\,\l)\,B(\vec b,\,\l)\ , \eel{probab}
where \(\r(\l)\) is positive and normalized. It is then proved, with mathematical rigor, that this outcome disagrees with the quantum mechanical expectation values. 

What this means is that the choice of the setting of a polarization filter \(\vec a\) by observer \(\cal A\)	cannot affect the action of polarization filter \(\vec b\) on the photon observed by observer \(\cal B\), if \(\vec a\) and \(\vec b\) are space-like separated. In quantum mechanics, Eq.~\eqn{probab} is violated, and by many researchers this violation was seen as an apparent violation of either the hidden variable concept, or of special relativity.

What does the CA theory tell us about Eq.~\eqn{probab}? Assume that the measurement devices \(\vec a\) and \(\vec b\), as well as the system that produces the quantum objects in question, such the photons in an EPR experiment, are all spacelike separated.

``Superdeterminism" definitely applies to this system, so that there is no such thing as ``free will", and we also do not share the revulsion often expressed against the idea of ``conspiracies". It is however correctly pointed out in numerous investigations, that this is not enough to escape from Bell's theorem. Let us sharpen the formulation of the problem.

The somewhat vague and misleading notion of ``free will" must be replaced by another demand that we think is more essential here: the \emph{freedom to choose our initial state}: a physical theory should provide meaningful predictions concerning the evolution laws \emph{regardless} our choice of the initial condition. Therefore, given the outcome of one experiment, we should be able to predict the outcome of another experiment, if we make just one modification, such as the setting of measuring device \(\vec a\). We keep the EPR photons, as well as detector \(\vec b\), in the same states as before. According to Bell, as is well-known, this leads to conflicts with Eq.~\eqn{probab}. This requirement now, therefore, does provide us with a fundamental problem. We will have to \emph{restrain the freedom to choose the initial condition.} How should this be done?
	\def\ua{\uparrow} \def\da{\downarrow} \def \la{\leftarrow}

For simplicity, we consider an experiment with entangled spin \(\half\) particles (of course, photons with spin 1  can be handled the same way, but the notation would be a little more clumsy).	
A particle polarized in the \,z \,direction is described by a wave function having as a factor a state \(|\ua\,\ket= ({1\atop 0})\) or \(|\da\,\ket=({0\atop 1})\). Particles polarized in the \,x \,direction have wave functions \(|\ra\,\ket=2^{-1/2}\,({1\atop 1})\) or \(|\la\,\ket=2^{-1/2}\,({1\atop -1})\). A pair of entangled particles has a wave function
	\be \fract 1 {\sqrt 2}(|\da\ua\,\ket-|\ua\da\,\ket)=\fract 1{\sqrt 2}(|\ra\,\la\ket-|\la\,\ra\ket=\fract 1 {\sqrt	 2}\pmatrix {0&1\cr -1&0}\ .		\eel{entangled}
	If we had photons, we would be working with polarization filters, but since here we have spin \(\half\) particles, we should work with what we will call Stern-Gerlach detectors, or spin detectors for short. They actually have the advantage that a particle can be detected completely, while its spin component in one pre-assigned direction can be established.

The difficulty describing Bell experiments in the CA theory is the following. If the spin detectors point in the \,z \,direction, each detected particle is split into the ``ontological" states \( |\ua\,\ket\) and \(|\da\,\ket\), and if the spin detector is orientated in another direction, say the \,x \,direction, then the splitting is in terms of the states \(|\ra\,\ket\) and \(|\la\,\ket\), which would then be the ontological states. The difficulty becomes clear immediately: the entangled particles, on their way to the spin detector, seem to know in advance which states are ontological. But how could these entangled photons ``know in advance" how to split up? How does the CA ``know" what the ontological states are? This, in a nut shell, is what many researchers denounce as ``conspiracy". If we try to avoid it, we end up with Eq.~\eqn{probab}, which cannot be right.

The correct answer to such paradoxes can always be found by realizing how the mapping goes between a CA and the quantum mechanical system that we are describing, and how a measurement takes place (see Section~\ref{axioms}). In this case, the answer       is to be found in the structure of the vacuum state. As stated in the previous subsection, all our particles are surrounded by a vacuum, and that state is highly entangled, containing \emph{spacelike} correlations in terms of the states of the CA. 

Now again consider the spin detector \(\vec a\), still spacelike separated from the place where the entangled particles originate. Modifying \(\vec a\) does not affect the wave function of the entangled particles, but it does affect how this wave function is to be expressed in terms of different ontological states. This we know for sure, because otherwise, Bell's inequalities would hold. How can we explain this?

The reason must be that the modified detector is surrounded by a vacuum whose correlations are incompatible with those of the vacuum surrounding the original detector. We know that this is possible because the vacuum has long-range, spacelike correlations. But there is another way to see this. Let us look at the CA states. After the experiment has been done, we have a definite ontological state, which we can extrapolate backwards in time. Besides the entangled state \eqn{entangled}, these particles can be in three other states. These other states however do not fit with the boundary condition at times before the particles originated, so these states do not agree with the original object surrounded by any acceptable vacuum with appropriate correlation functions. From this it follows that the vacuum correlations at the moment that the entangled particles were formed, were very special, and they depend in a crucial manner on the spin detectors at some spacelike distance away. Indeed, this could be called ``conspiracy", but it is conspiracy of an acceptable nature: no signals can be sent faster than light. This conspiracy arises from spacelike correlations that are quite familiar in quantum field theory; the correlations are expressed by Green functions with points that are spacelike separated. Like the propagators in QFT, such spacelike correlations do not vanish. The physical reason is simple: the fluctuations arise from oscillations in a common past.

We note that, therefore, modifying the spin detector \(\vec a\) without modifying the wave function of the entangled particles is allowed, but modifying it without modifying the ontological interpretation of these particles is not allowed. Apparently, we can't modify the initial state anyway we like: the definition of the vacuum state does not allow this. This also implies that our modified notion of ``free will" as formulated above is not valid.

We observe that non-locality of the CA itself is never needed for this argument. A cellular automaton can be local in the sense that the evolution happens in small discrete steps in the time coordinate, where every cell is updated in a way depending only on the contents of neighboring cells. 

\subsection{Conspiracy}\lab{conspiracy}

It is not so easy to dismiss the above observations as acts of ``conspiracy" by Nature's laws. To illustrate this, let us take an example of  ``conspiracy" from number theory. Consider a very large prime number \(P\). Consider another, even larger number \(Q\) defined by
	\be Q\equiv 2^{P-1}-1\ .\eel{twopower}
What are the odds the \(Q\) can be divided by \(P\)? A beginning student might argue that, if \(P\) is hundreds of digits long,  the odds against that are tremendous. Simple number theory however says that \(Q\) is \emph{always} dividable by \(P\), provided \(P\) is a prime. If \(P\)is not a prime, it nearly never is. The argument goes by simple counting, but one has to count all the way to \(P\). Is this a conspiracy? Perhaps, but it also is a fact.

In our theory, all conspiracy goes away as soon as we realize that the most efficient way to describe a large CA is indeed by replacing its classical evolution law by a Schr\"odinger equation, which is known to yield exactly the correct results. The CA itself may seem to be conspirational, if it is sufficiently non-trivial due to interactions. These interactions also turn the CA into a universal computer, whose actions cannot be compressed as efficiently as a quantum field theory. 

\subsection{Quantum computers}\lab{qpc.sec}

It is sometimes put forward that our findings imply that our quantum system should be able to represent a quantum computer, whereas the CA, though very powerful, can at best simulate a classical computer. Quantum computers are expected to solve non-polynomial problems, which a classical computer cannot. Here, our response is that, a quantum computer can only outperform a classical one if the latter works perfectly. The quantum systems generated by CAs will in general be ugly, containing various types of interactions. These interactions may well cause limitations of a fundamental type to the performance of any quantum computer.

\subsection{Classical information loss}\lab{infoloss} Before phrasing the axioms of the CA interpretation, an important generalization must be mentioned. It is conceivable that the classical automaton is \emph{not} time-reversible. This means that, given the configuration at time \(t=t_1\), there could exist more than one different configurations at \(t=t_0<t_1\) that evolve into the given one, or there could be none at all (``Garden of Eden"). At first sight, this might be considered unworkable, but in principle, such a CA can be used. What has to be done is that the CA states must be assembled in \emph{equivalence classes}: two states are equivalent if, at some time in the foreseeable future, they both evolve into the same state. By construction then, every equivalence class evolves into exactly one equivalence class in the future, and it has one equivalence class in its past (that class could be the empty class, but once classes have grown to be large enough, this case might be negligible).

These equivalence classes are very similar to gauge equivalence classes: it is impossible to distinguish experimentally which element of the equivalence class we have, because this information disappears after a finite amount of time.

We now associate one normalized quantum basis element to each equivalence class. Our quantum Hilbert space then must be the Hilbert space spanned by these basis elements. 

A notable feature of this Hilbert space is that it might be much smaller than if no information loss took place. It could even be that, for a given volume in 3-space, the equivalence classes can be associated with the structures at the surface only, in case all information in the bulk becomes unstable. This could be an interpretation of ``holography" in quantum gravity and black hole physics.

An advantage of this scenario could be that the association between quantum states and ontological states may become non-local to a large extent, which could well explain the apparent contradictions between the simple-minded time-reversible CA theories and the attempted interpretation of Bell experiments in quantum mechanics as we know it. We would have a peculiar, but desirable feature: Quantum theory is local in the sense that commutators vanish outside the light-cone, and the classical CA theory is local as well, but the mapping between the quantum states and the CA equivalence classes is an apparently non-local one. This could be regarded as a possible loophole to reduce apparent contradictions with Bell's theorem, but this remains to be seen. As yet, the mapping we considered does not seem to be `sufficiently' non local, and it is our suspicion that the occurence of \emph{spacelike correlations in the vacuum} (subsection~\ref{vacuum.sec}), perhaps in combination with the \emph{ambiguity of the wave function}, as emphasized in subsections~\ref{CAinterpr} and \ref{Bell.sec}, is a more elegant loophole.

It is noteworthy that the superstring theory that we worked out in this paper is time-reversible, so information loss was not put in. On the other hand, we do know that string theories are associated with black holes where we do have holography. This is why we put information loss on hold in this paper. It just gives one enough room to speculate that more complete models to be constructed in the future may handle the non-locality issues in a more complete manner.

\newsecl{Axioms of the CA interpretation}{axioms}

In view of the above, we can now attempt to phrase the essentials of the CA interpretation as succinctly as possible. We expect that these might be sharpened when the theory becomes more mature. Let us start from the quantum system we wish to describe.
\bi{$i$} We consider quantum theories that obey all postulates of standard quantum mechanics in agreement with the Copenhagen interpretation, except that we put an important constraint on what can \emph{eventually} be detected, see below, under $(iv)$.
\itm{$ii$} In many interesting theories, a special basis in Hilbert space can be found that we will call the ontological or CA basis. This is an orthonormal set \(\{|\vec Q(t)\ket\}\) that spans all of Hilbert space. Only these basis elements, \(|\vec Q(t)\ket\),  have the property that the evolution operator over specially chosen time steps simply \emph{permutes} these states. They describe all ontic states. We may describe them as depending on a time variable \(t\). The phase angles of these states are physically meaningless.
\itm{$iii$} All other states \(|\j\ket\) in Hilbert space describe probabilistic distributions,\\ \(\r(\vec Q)=|\bra \vec Q|\j\ket|^2\).
Again, the phases of the CA states are meaningless. However, if we fix the phases for the CA states, we can define uniquely the evolution operators \(U(t_1,t_2)\), and write these as \(e^{-iH(t_2-t_1)}\). 
\ei At this stage, it is emphasized that a generic quantum state \(|\j\ket\) may have any set of phases in its inner products \(\bra\vec Q|\j\ket\). Changing these phases does not have any effect on the physics. This means that many different wave functions can be used to describe the same reality. In a more conventional basis (to be referred to as the SM basis), the hamiltonian for these different wave functions will look totally different since the phases enter into the expressions for \(H\). Note that it would not be a good idea to choose all phase factors  of the \(|\vec Q\ket\) states to be \(+1\), since we wish to use the freedom to perform superpositions. 

 The fact that (time-dependent) unitary transformations may exist as invariances will be an important feature of the quantum theory, but such invariances may be difficult to identify in the real world.
\bi{$iv$} Only the projection postulate changes: one cannot project a quantum state \(|\j\ket\) onto, say, an eigenstate \(|\chi\ket\) of an arbitrary operator, to measure \(\bra\chi|\j\ket\). To measure something, we must make a model of the measuring apparatus that eventually produces a macroscopic effect. This macroscopic effect can always be represented by the ontological states \(|\vec Q(t)\ket\), so that the state of the detector can be measured by projecting \(|\j\ket\) on these CA states.\ei
This presupposes that such detectors exist in the given CA model, which is far from obvious, but we know of course that they exist in the real world. This requires the following assumption:
\bi{$v$} The ontological states \(|\vec Q\ket\), equipped with arbitrary phase angles, are assumed to be locally dense in Hilbert space. This means that in any restricted region of space-time, one can approach any given quantum state \(|\j\ket\) with a CA state; orthonormality is restored by differences outside that region. \ei 
The accuracy of the approximation, or the question exactly \emph{how} dense this subset is, may depend on the volume, this will be worth further investigation.
\bi{$vi$} The complete vacuum is not an ontological state, but a very delicate superposition. Locally it can be approached by CA states, see $(v)$. Conversely, a vast majority of the CA states can never be reaized in the real world because they represent improbable amounts of energy compressed in very small regions. 
\itm{$vii$} Conversely, we can start with a cellular automaton (CA) that evolves classically. This evolution can then be represented by an evolution operator. Standard quantum mechanics can be recovered from that by performing unitary transformations and constructing an appropriate hamiltonian. \ei 
This may actually lead to several quantum models representing the same CA, for instance by redefining the phases of the states \(|\vec Q\ket\), or by adding conserved quantities to the hamiltonian.
\bi{$viii$} Because of the non observability of the phase factors of the CA states, each probability distribution \(\r(\vec Q)\) of the CA states represents a very large class of apparently different wave functions.
\ei
This implies that replacement of a wave function by an element of the same class has no physical effect but may be required for understanding of how Bell's theorem is evaded.

The work presented in this paper suggests that mappings of the kind used here do exist; in fact, they may seem to be  rather easy to construct except that in most cases, either the CA or the quantum system are too far remote from the desired theories. It is for instance difficult to impose rotation invariance and even more difficult to impose Lorentz invariance, let alone general coordinate invariance. It is these latter requirements where we now have made progress.

\newsecl{Discussion and conclusions}{disconc}

We think that something very important has been discovered:
\begin{quote} Both the bosonic string theory and superstring theory can be reformulated in terms of a special basis of states, defined on a space-time lattice, with lattice length \(2\pi\sqrt{\a'}\).  The evolution equations on this lattice are classical. This allows for a CA interpretation of superstring theory.\end{quote}
Although the lattice is defined in Minkowski space, Lorentz invariance is not generally guaranteed. This is where we can use established knowledge of (super)string theory to conclude that space-time must be 26 or 10 dimensional. \emph{Discrete} Lorentz invariance may be easier to establish, although there are a few considerations that have to be taken into account.
Discrete Lorentz invariance would be evident if the superstring constraint \eqn{boseconstr} would be replaced by constraint \eqn{Aconstraint} of the discrete theory, but they do not seem to be the same.

There are a number of issues that we briefly address in this Section.

\subsection{Classical vs. quantum mechanics}\label{classquant}
	The main theme of this paper is that a mapping is found between a classical (CA) description of a superstring, and the usual quantum mechanical one. Here is what we have:  
\bi{-} The independent degrees of freedom of strings and superstrings are the transversal string coordinates and spinorial fields. On the world sheet of the string, they all obey strictly linear field equations, such as:
	\be q^\m(x,t+\d)	=q^\m(x-\d,t)+q^\m(x+\d,t)-q^\m(x,t-\d)\ ,\eel{latticeequ}
where \(\d\) can be any fixed number.
This simply follows from the fact that there are left-movers and right-movers. Because of this, these  degrees of  freedom can be decomposed into lattice degrees of  freedom without loosing any generality. The lattice has light like coordinates with lattice length \(\d\), but at any stage, we can choose \(\d\) to be something else, so it should be easy at a later stage to take the continuum limit. 
\itm{-} By splitting up the transverse bosonic string degrees of freedom \(q^a\) \((a=1,\cdots, D-2)\) into their integer parts and their fractional parts (a precise description of this splitting being a little more complex than this, see Ref.~\cite{GtH2}), we find that also space-time itself can be limited to being a lattice. The fractional parts actually also can be transformed to being a lattice, simply by Fourier transforming: they form the dual lattice.
\itm{-} Only the commutators between the lattice and the dual lattice variables are non-vanishing. The lattice degrees of freedom all commute with one another. Also the dual lattice degrees of freedom all commute with one another. Note that the dual lattice and the lattice itself describe the same physics. 
\itm{-} Because all variables on a lattice commute, a basis of states exists such that, if a wave function starts off as having only 1s and 0s, it will always be restricted to 1s and 0s. This means that a formalism is found in terms of which states that are ``certainly true" evolve into other states that are ``certainly true". There will never be quantum superpositions. This basis is referred to as the `ontological basis'.
\itm{-} If the (super)string \emph{interactions} are described as exchanges of legs when strings intersect, then it seems that these can also be described as ``certainties": there still will be no quantum superpositions in the CA description.  This however may only be true if the string interaction constant has a fixed value.
\ei For (super) string theory this may be an interesting observation: not all string theories are equal; these will be special, having more predictable properties.
\bi{-} Quantum superpositions will occur as soon as we try to describe the string modes in terms of a quantum field theory. This is just the passage towards another quantum basis. Only in terms of such a basis, indeed any of the basis choices usually employed in quantum field theory, or quantum theory in general, interference phenomena are perceived as real, and this is why we do encounter quantum phenomena in ordinary physics.
\ei
Most of our \emph{symmetry transformations} only exist in the quantum description, not on the ontological basis. Rotations in the transverse dimensions are continuous only in terms of the real variables \(q^a(x,t)\), not after we made the transition to the target space lattice variables, where only discrete crystal transformations can be done. Lorentz transformations are even more subtle; they also involve the world sheet lattice in a complicated way.

The philosophy used here is often attacked by using Bell's inequalities\cite{Bell}---\cite{CK} applied to some Gedanken experiment, or some similar ``quantum" arguments. The apparent \emph{no-go} arguments have to be considered carefully. They are very important to put things in perspective, and in general they are employed to conclude that there are still essential elements missing in the picture we have. In the present version of this paper, we addressed the associated issues in Sections \ref{CAinterpr.sec}---\ref{axioms}.

We emphasize that the mathematical expressions we found corroborate our views.

\subsection{The world sheet lattice}\label{pplus}
	Our theory does away with all distance scales shorter than the lattice length \(a_\ell\) in a very effective way.	
The string degrees of freedom are split up in their integer parts and fractional parts (in units where \(a_\ell=1\)). The fractional parts are redundant because they merely refer to operators acting on the integer parts.

In conventional formulations of string theory, one often picks the gauge choice
	\be \pa_+q^+=\half(\pa_x+\pa_t)q^+=a_L^+=p^+\ , \eel{pplusgauge}
where \(p^+\) is the total light-cone momentum, a conserved quantity. One then gets
	\be \pa_+q^-=\half(\pa_x+\pa_t)q^-=a_L^-={1\over p^+}\bigg((a_L^{\mathrm{tr}})^2+C\bigg)\ ,\eel{aminus}
where \(C\), the mass-squared of the lowest state, or intercept, is either tachyonic or it vanishes.

For the continuum theory, the \(p^+\) dependence is of no consequence, as it is dictated by a trivial component of the Lorentz transformations. One can choose the Lorentz frame where it is one. In the space-time lattice, however, it is quite non-trivial, since both the left movers \(A^L\) and the right movers \(A^R\) are now constrained to be integers. If we limit ourselves to the discrete Lorentz group \(O(D-1,1,\mathbb Z)\), then \(A^\pm_{L,R}\) are all integer, but if the sequence \(A^+_L(x+t)\) takes on different integer values, then just setting these all equal to one, as in the gauge choice \eqn{pplusgauge}, at first sight seems not to be allowed. 

If at one spot, \(A_L^+\) takes a value \(N\), one might consider adding \(N-1\) dummy lattice sites between two consecutive string joints there, but then soon the question will arise how to assign the other \(A_{L,R}^\pm\) values at these intermediate points. To answer this question, we believe it will be best to go to the continuum limit on the world sheet, but the logic of that must be further investigated. 
Our point in this subsection is, that at all stages one has the freedom to choose the gauge, and choosing the longitudinal components of the physical degrees of freedom serves no other purpose than to establish Lorentz invariance. We actually have to do this in the continuum theory differently  from the lattice theory, as the next subsection shows.

\subsection{The string constraints and Lorentz invariance}\label{strconstr}
	At first sight, the superstring constraints \eqn{worldgauge}---\eqn{fermconstr}, seem to be so similar to the constraints \eqn{Aconstraint} and \eqn{Azerobound}, that one might think they can be mapped easily. This however, does not seem to be the case. We have the problem of the superfluous world sheet lattice coordinates, as displayed in the previous subsection. But there is also the problem that on the lattice only a limited number of Fourier components occur, while the continuum has infinitely many.
	
	In the continuum, the Lorentz group that can be constructed closes algebraically, due to the special linear properties of the commutation relations of the string excitation modes, though only if \(D=26\) or \(10\). On a finite lattice, the commutation rules among the excited modes become more complicated and the Lorentz group does not close. This is linked to the fact that the lattice requires a preferred gauge, which no longer holds after a Lorentz transformation. We believe that this is a consequence of the Lorentz group here being discrete rather than continuous. Continuous Lorentz invariance is not self-evident, and indeed may often fail. Indeed, all we have at present is the conjecture that continuous Lorentz invariance emerges spontaneously if we build in discrete Lorentz invariance. Our reasons for this conjecture is that we do have emergent invariance under continuous \(O(D-2,\mathbb R)\) rotations, which, together with invariance under the discrete Lorentz group \(O(D-1,1,\mathbb Z)\), should generate continuous Lorentz invariance.
	
	Regarding this latter argument, one may now wonder why this can go wrong in the cases that \(D\ne 26\) and \(\ne 10\). We then must note that there may be subtle features that jeopardize discrete Lorentz invariance. There is a state counting problem.
	
	Suppose the transverse degrees of freedom \(A^{\mathrm{tr}}_{L,R}\) are all given. This means that the longitudinal components are fixed by their products, \(A^+_L\,A^-_L\) and  \(A^+_R\,A^-_R\), and the fact that all these variables are integer. This, however, does not fix their values completely, so we must choose the world sheet gauge more precisely, or else accept that different states are described by the same transverse data. Either we under-count or we over-count states this way. Exactly how to do this right is still an open question. Perhaps details of the lattice are important to resolve this problem: are all sites occupied? Do certain even or odd positions stay empty? Do we have to consider different space-time ``crystals"? One is tempted to suspect that this is a source where the dimensionality of the lattice plays an important role.
	
\subsection{Why fermions? Why supersymmetry?}\label{fermisuper}
Fermions are added to the discrete deterministic theory by inserting \(D-2\) Boolean degrees of freedom to every lattice site on the world sheet. In the time evolution, these Boolean degrees of freedom are passed on as left- and right movers just as in a cellular automaton. It was easy to turn these into genuine fermions by means of a Jordan Wigner transformation\cite{JW}. In the bulk theory, each fermionic mode does not seem to be directly linked to any of the transverse directions on the lattice, so their spin structure stays obscure. This changes only when the constraints are considered, which means that their properties under rotation become clear only when Lorentz transformations are considered. Galilean transformations may suffice, so this means that the spin rotation properties are revealed as soon as these particles move, in target space. Since both fermions and bosons are massless and non-interacting, having exactly \(D-2\) fermions automatically leads to supersymmetry, which is also necessary to remove the longitudinal modes.

In our formalism it is as yet hard to see why the presence of these fermions affect the number \(D\) of space-time dimensions. An observation to be made is that the state counting problem is intimately tied to the question how exactly the space-time lattice should be occupied. Are there empty sites? In the space of integers, one finds that sometimes it makes a difference whether world sheet lattice sites are characterized by even or odd integers. This may give rise to an interesting speculation concerning the fermionic degrees of freedom. Maybe the Boolean variable \(\s(x,t)\) indicates whether the lattice site occupied at \((x,t)\) is even or odd. To find out more about this, the state counting problem may have to be addressed, and this was as yet too complicated to consider here.

\subsection{The lattice length in space-time}\label{length}
On the world sheet lattice we saw that the even sites (where \(x+t\) is even) are detached from the odd sites (\(x+t\) odd). Also in the target space (space-time) lattice, it may be important to distinguish even and odd integers. If, for instance 
	\be A_L^+A_L^-=\sum_{a=1}^{D-2}(A_L^a)^2\ ,\nm\ee
(and similarly for the right-movers), we see that \(A_L^+\) and/or \(A_L^-\) are even or odd depending on the number of even or odd values of \(A_L^a\), modulo 4. Note that if \(A^+_L\) is odd, then so is \(A^-_L\). So it may happen that the periodicity of the lattice is twice (or more times?) the minimal integer number occurring there. We normalize the minimal integer on this lattice as the number \(a_\ell\), so the lattice period may be a multiple of this. In our calculations, we have been using units where  \(a_\ell=1,\ c=1\), and \(h=2\pi\hbar=1\). In these units,
	\be a_\ell=h\sqrt{\a'}=2\pi\hbar\sqrt{\a'}\ . \eel{alphaprime}
The string coupling constant \(g_s\) is also one in our theory, since it describes the exchange operation which is indeed unity if we also insist that the interaction respects determinism. Since \(g_s\) also fixes the strength of the gravitational force, we get that \(\a'\) must be close to the Planck length squared, and our space-time lattice is close to the Planck length\fn{Remember that the actual physical value for the Planck length may depend on the compactification scale of the extra dimensions.} itself.

\section*{Acknowledgements} The author thanks in particular M. Porter, R. Maimon and P. Shor, for their constructive criticism.


\end{document}